\documentclass[emulateapj,psfig,apjfonts]{emulateapj}
\submitted{{\it Accepted for publication in PASP}}

\usepackage{graphicx}
\usepackage{natbib}

\begin{document}

\title{The $\chi$ Factor: \\ Determining the Strength of Activity in Low Mass Dwarfs}

\author{Lucianne M. Walkowicz}
\affil{Astronomy Department, University of Washington, Box 351580, Seattle, WA 98195-1580}
\email{lucianne@astro.washington.edu}

\author{Suzanne L. Hawley}
\affil{Astronomy Department, University of Washington, Box 351580, Seattle, WA 98195-1580}
\email{slh@astro.washington.edu}

\author{Andrew A. West}
\affil{Astronomy Department, University of Washington, Box 351580, Seattle, WA 98195-1580}
\email{west@astro.washington.edu}

\begin{abstract}
We describe a new, distance-independent method for calculating the magnetic activity strength in low mass dwarfs, L$_{\rm{H}\alpha}$/L$_{\rm{bol}}$. Using a well-observed sample of nearby stars and cool standards spanning spectral type M0.5 to L0, we compute ``$\chi$'', the ratio between the continuum flux near H$\alpha$ and the bolometric flux, f$_{\lambda6560}$/f$_{\rm{bol}}$. This ratio may be multiplied by the measured equivalent width of the H$\alpha$ emission line to yield L$_{\rm{H}\alpha}$/L$_{\rm{bol}}$. We provide $\chi$ values for all objects in our sample, as well as fits to $\chi$ as a function of color and average values by spectral type. This method was used by \citet{we04} to examine trends in magnetic activity strength in low mass stars.
\end{abstract}


\section{Introduction}
Low mass stars are the most populous objects in our Galaxy, comprising roughly 70$\%$ of the stars in the Solar Neighborhood and almost half of the Galactic stellar mass [\citet{he97}]. Until the last decade, the low luminosities of these stars prevented their study by large scale spectroscopic surveys. However, efforts such as the Palomar/MSU Nearby Star Spectroscopic Survey [\citet{re95a}, \citet{ha96}, \citet{gi02}, \citet{re02}], 2MASS followup [\citet{gi00b};  \citet{gi00a}], and NLTT followup [\citet{re04}, \citet{le03}] have characterized global properties of nearby M dwarfs with samples numbering $\gtrsim$ 1000 objects. Recently, the Sloan Digital Sky Survey (SDSS) has made it possible to study spectroscopic samples approaching 10,000 low mass stars [\citet{we04}]. 

In particular, these surveys have provided insight into the magnetic behavior of low mass stars. There are many diagnostics by which a star may be deemed magnetically active, including X-ray and FUV emission lines, but in optical wavelengths, H$\alpha$ is the most readily available signature of activity. Although H$\alpha$ in either emission or absorption may indicate the presence of a chromosphere, most large-scale optical studies have used the more restrictive definition of H$\alpha$ in emission. This practice is due to the difficulty in breaking the degeneracy between intermediate activity and truly weak chromospheres, both of which present as weak H$\alpha$ absorption (see \citet{gia82} and \citet{gi02} for a discussion of this phenomenon). In this paper we use ``active'' to mean the presence of H$\alpha$ in emission.

Increasingly large studies of active M dwarfs have built on one another to create our current picture of activity at the bottom of the main sequence. Originally, \citet{jo74} showed a rise in the fraction of active M dwarfs towards later spectral types, with 100$\%$ of M5 stars showing activity. The PMSU investigation of \citet{ha96} extended the trend with much better statistics, and found that the fraction of active stars peaked at  $\sim$60$\%$ at spectral type M5.5. \citet{gi00b} observed a sample of later type stars from 2MASS, and found that the incidence of activity rose to 100$\%$ at type M7, but declined at later types. \citet{bu00} added additional late M and L dwarfs to the sample and confirmed that the active fraction decreases past type $\sim$M7 - M8. Most recently, \citet{we04} used a much larger sample of low mass dwarfs from SDSS to further refine the trend: these data reveal a curiously elegant bell curve distribution for the incidence of magnetic activity in low mass stars. The fraction peaks at $\sim$73$\%$ of M8 stars being active.

Several of these studies consider the strength of activity, which is quantified as the ratio of the luminosity in H${\alpha}$ to the bolometric luminosity. The activity strength remains relatively constant (though with large scatter) at log(L$_{\rm{H}\alpha}$/L$_{\rm{bol}}$) $\sim$-3.7 through the early to mid-  M spectral types, and appears to decline at types later than M6, while the active fraction is still rising. The increasing fraction of active stars may be an age effect (lower mass stars retain their activity longer) on the rising branch (M0 - M8).The activity strength results indicate that some damping mechanism apparently sets in around type M6. At types later than M8, this damping mechanism [possibly related to the atmospheric temperature, see \citet{mo02}] overwhelms the age effect and results in fewer active dwarfs at lower atmospheric temperature. The activity fraction and activity strength results, taken together, provide important constraints on models of magnetic dynamos and magnetic field production in M and L dwarfs, particularly as these objects span the mass range where the stars become fully convective and develop degenerate cores.

Quantifying the strength of activity as the ratio between H${\alpha}$ luminosity and bolometric luminosity, while informative, is prone to systematic error. Determining luminosities depends on having well-known distances, and as samples get larger (e.g. from SDSS) the stars are typically well beyond the immediate Solar Neighborhood (i.e. $>$ 25pc from the Sun). Thus, precise trigonometric parallax measurements are generally not available as in previous local surveys. In principle, one should be able to calculate the ratio L$_{\rm{H}\alpha}$/L$_{\rm{bol}}$ and expect the distance dependencies to cancel. However, in order to get L$_{\rm{bol}}$, one must have determined M$_{\rm{bol}}$, which relies on knowing M$_V$, having excellent color measurements, and a reliable bolometric correction. M$_V$ generally comes from distance estimates based on spectroscopic and/or photometric parallax determinations, which are not always reliable or in agreement, due primarily to a paucity of calibration data at the bottom of the main sequence. Calculating L$_{\rm{H}\alpha}$, on the other hand, relies on knowing the flux in the continuum near H$\alpha$. This is often obtained from the photometric colors [\citet{re95b}]. Alternatively, it may be measured directly from spectrophotometrically calibrated data. Neither method is necessarily congruous with that used to obtain L$_{\rm{bol}}$. Thus, in practice, the ratio usually contains two distances, typically determined in different ways, leading to possible systematic error. 

An alternative method is to determine the ratio of the flux in the continuum near H${\alpha}$ and the bolometric flux. This ratio, which we call $\chi$, can then be multiplied by the equivalent width in H$\alpha$ for a given star to give the ratio L$_{\rm{H}\alpha}$/L$_{\rm{bol}}$. If one has excellent quality data, the ratio f$_{\lambda6560}$/f$_{\rm{bol}}$ may be calculated for each of the stars observed. In this paper, we use a well-observed sample of nearby M dwarfs and early L dwarfs to determine $\chi$ as a function of several colors (I$_C$-K$_s$, V-I$_C$, $r - i$, $i - z$) and of spectral type, in effect providing an improved ``bolometric correction'' for low mass dwarfs.

\section{The Sample and Method}

We drew our sample of calibrating stars for early spectral types (M0.5 to M6.5) from the nearby 8 parsec sample [\citet{re97}; \citet{nlds}, Appendix A; \citet{re95a}, \citet{ha96}] and used a sample of cool standard stars from 2MASS for later spectral types (M4.5 to L0) [\citet{gi00a}; \citet{ki99}; \citet{ki00}; \citet{re99}]. The overlap of the two samples at midrange spectral types (M4.5-M6.5) ensured that there were no systematic differences in f$_{\lambda6560}$/f$_{\rm{bol}}$ for the two samples. The 8 parsec sample has been well-characterized photometrically (in the UBVR$_C$I$_C$ Johnson-Cousins filters as well as near-infrared J, H, and K filters) and spectroscopically. Spectra for the 8 parsec sample were largely obtained using the spectrograph on the Palomar 60 inch telescope with a 600 l/mm grating, blazed at $\lambda$ 6500, giving a dispersion of 1.5 $\mbox{\AA}$ pix$^{-1}$ and coverage from $\sim$6200 to 7400 $\mbox{\AA}$. A 1 arcsec slit was used, giving a resolution element of $\sim$2.5 pixels. Most stars fainter than V=16 were observed using the double spectrograph on the Palomar 200 inch Hale telescope using the 6800 dichroic as a beam splitter, with the blue camera set to cover $\lambda$6200 - 6800 at a resolution of $\sim$0.55 $\mbox{\AA}$ pix$^{-1}$ and the red camera covering $\lambda$ 6900 - 7500 at $\sim$0.6 $\mbox{\AA}$ pix$^{-1}$. A 1 arcsec slit gives resolution of $\sim$3 pixels. Many of the 2MASS standards were observed using the low-resolution imaging spectrograph (LRIS) on the Keck I 10m telescope on Mauna Kea, Hawaii. The 1200 l/mm grating, centered at $\lambda$ = 6800 $\mbox{\AA}$, yielded coverage from $\lambda$6120 to 7500 at 0.5 $\mbox{\AA}$/pixel. With a 1 arcsec slit, this gave a resolution of 1.5 $\mbox{\AA}$. Visible photometry for the sample of late spectral types was not readily available, so I$_C$ magnitudes were found by convolving the spectra with a Cousins I$_C$ filter response curve. Infrared magnitudes for the late type objects were obtained from the 2MASS database. All spectra for the objects in our sample were spectrophotometrically calibrated in the region near H$\alpha$.

\begin{figure}
\begin{center}
\includegraphics[width=0.7\textwidth]{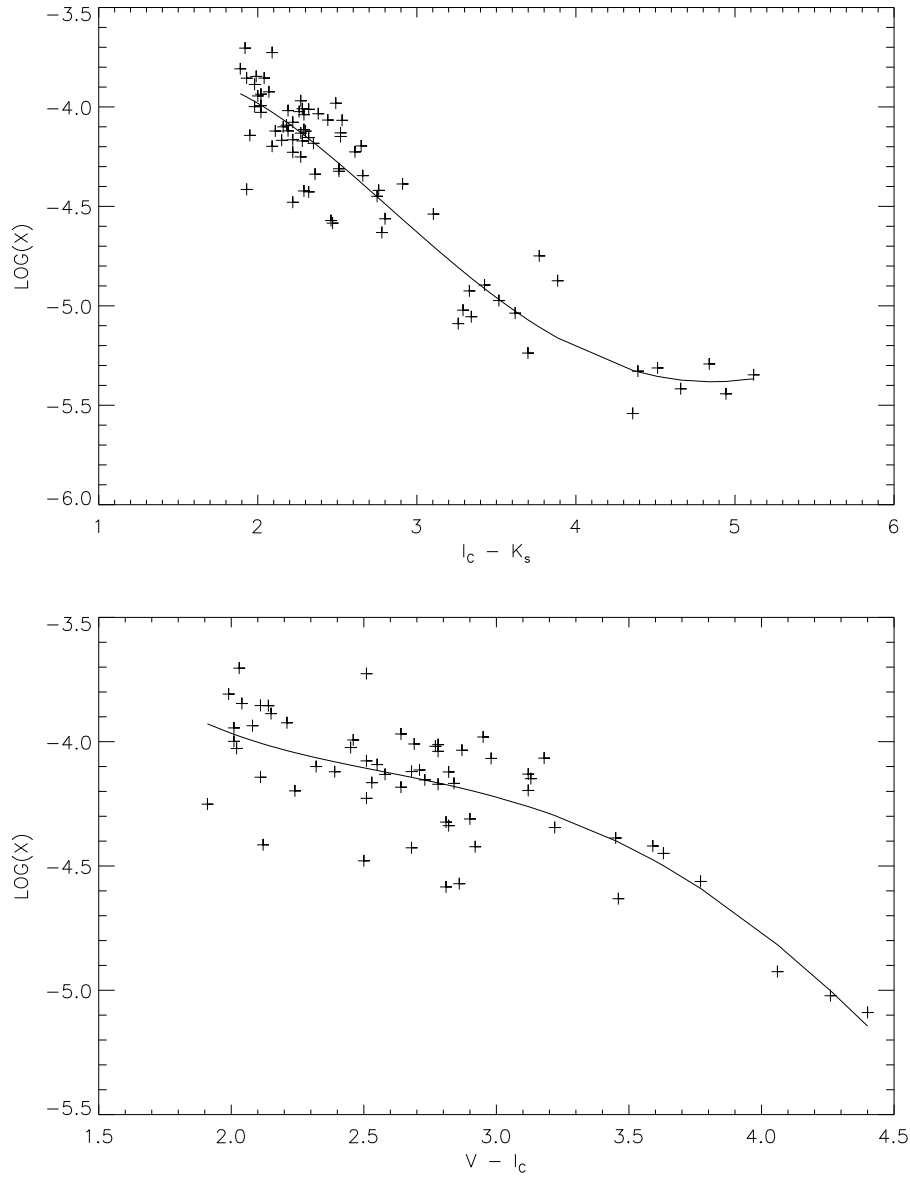}
\caption{$\log(\chi)$ versus  I$_C-K_s$ (upper) and V$-I_C$ (lower). Data are shown as individual crosses; the fit given in the text is the solid line in each panel.}
\label{fig1}
\end{center}
\end{figure}

The method to find $\chi$ is straightforward. Photometric data for the 8 parsec sample were used with the bolometric corrections of \citet{le96} to find apparent bolometric magnitudes for the sample. The bolometric magnitudes, m$_{\rm{bol}}$, were then transformed to apparent bolometric fluxes, f$_{\rm{bol}}$. The values of the apparent continuum flux near H${\alpha}$, f$_{\lambda6560}$, defined as the mean between 6555$\mbox{\AA}$ and 6560$\mbox{\AA}$, were measured directly from the calibrated spectra. For the late-type stars, the spectra were convolved with a Cousins I$_C$ filter response, and K$_{s}$ (the short K filter used by 2MASS) was transformed to K$_{UKIRT}$ (using relations from \citet{ca01}) for use in the bolometric correction relations. Bolometric corrections were calculated using the \citet{le00} relation in I$_C$-K$_{UKIRT}$, which is more reliable for late spectral types. Bolometric fluxes were computed as for the earlier spectral types, and continuum fluxes at H$\alpha$ were measured directly from the spectra. Combining the samples, we obtain $\chi$ = f$_{\lambda6560}$/f$_{\rm{bol}}$ over the entire spectral range. The mean values of $\chi$ by spectral type are shown in Table \ref{tbl-1}, while specific values for each object in the sample are given in Table \ref{tbl-2}. Figure \ref{fig1} shows the data and fit to $\chi$ in I$_C$ - K$_s$ and V - I$_C$, and Equations 1 and 2 provide the fits explicitly. Comparison in the overlap region of M4.5 V to M6.5 V shows no difference in the behavior of $\chi$ for the two samples.  

\begin{figure}
\begin{center}
\includegraphics[width=0.7\textwidth]{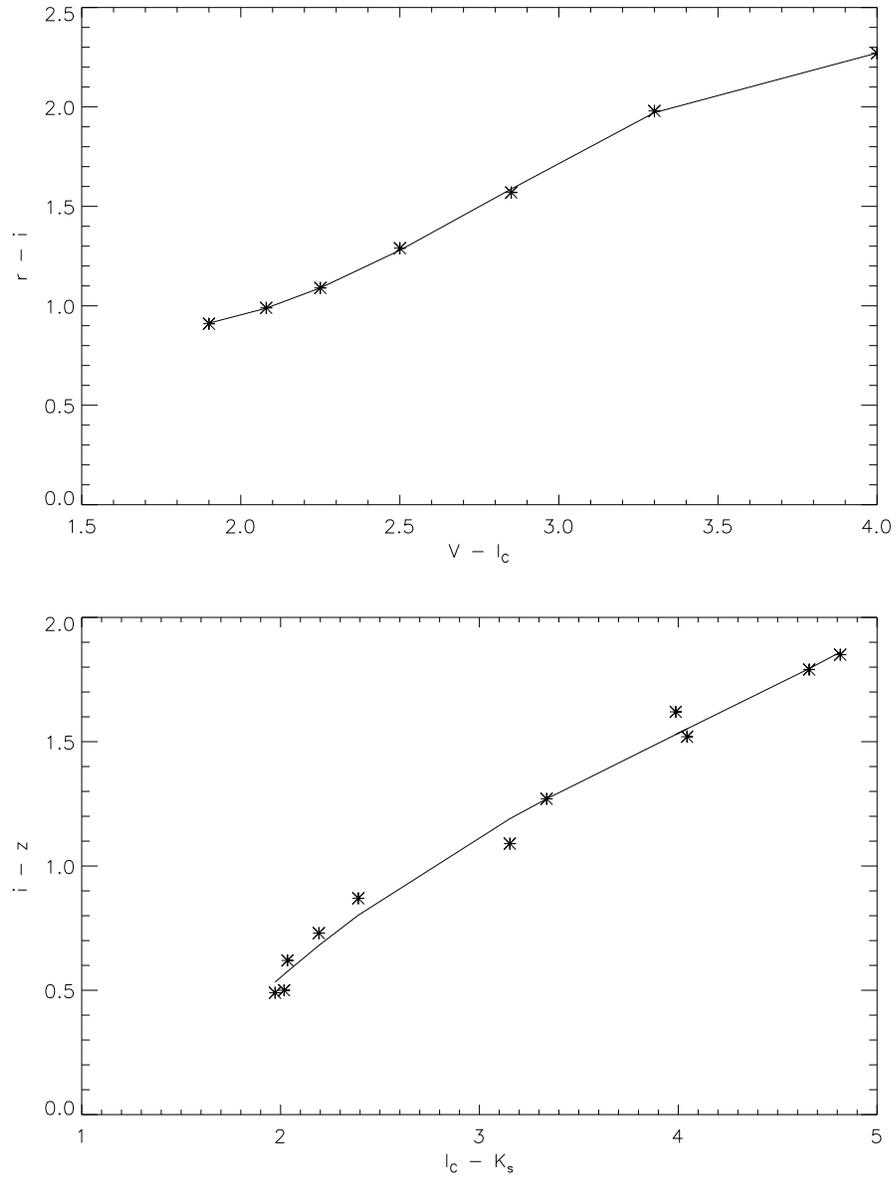}
\caption{Data and fits between SDSS and Cousins-2MASS colors: $r - i$ versus V$-I_C$ (upper) and $i - z$ versus I$_C-K_s$ (lower).}
\label{fig2}
\end{center}
\end{figure}

\begin{figure}
\begin{center}
\includegraphics[width=0.7\textwidth]{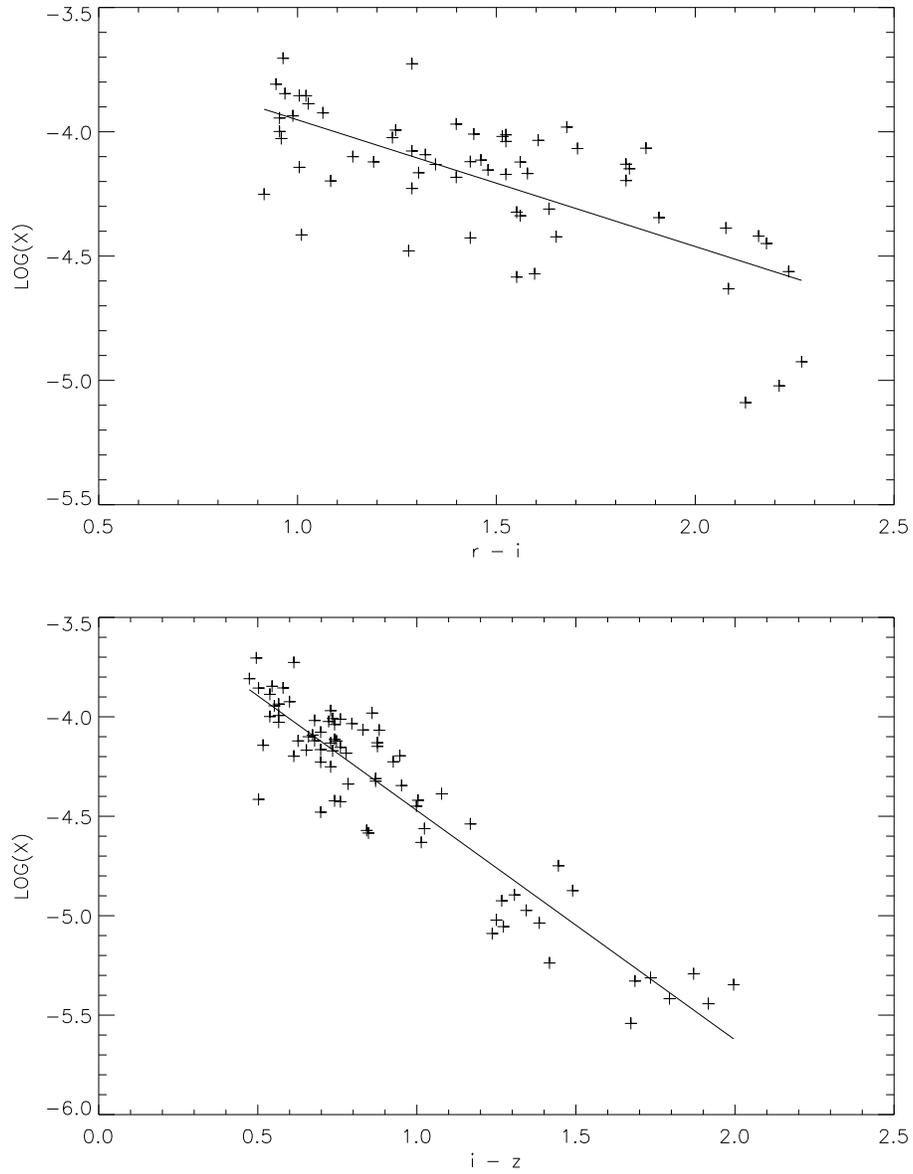}
\caption{$\log(\chi)$ versus $r - i$ (upper) and $i - z$ (lower). Original colors transformed to SDSS using fits to average color by spectral type.}
\label{fig3}
\end{center}
\end{figure}

We are particularly interested in being able to use $\chi$ with our large sample of SDSS spectra. Unfortunately, transformations between the SDSS filter system and other common astronomical systems do not extend into the very red color range we require here. To address this problem, we computed fits between average colors by spectral type in traditional and SDSS filters. These fits were used to transform the colors for our sample onto the SDSS system. In the following section, Figure \ref{fig2} shows the color transformations used, while Equations 3 and 4 provide the relations. Figure \ref{fig3} shows the $\chi$ results for the SDSS $r - i$ and $i - z$ colors, and the fits are given in Equations 5 and 6. 

\begin{figure}
\begin{center}
\includegraphics[width=0.7\textwidth]{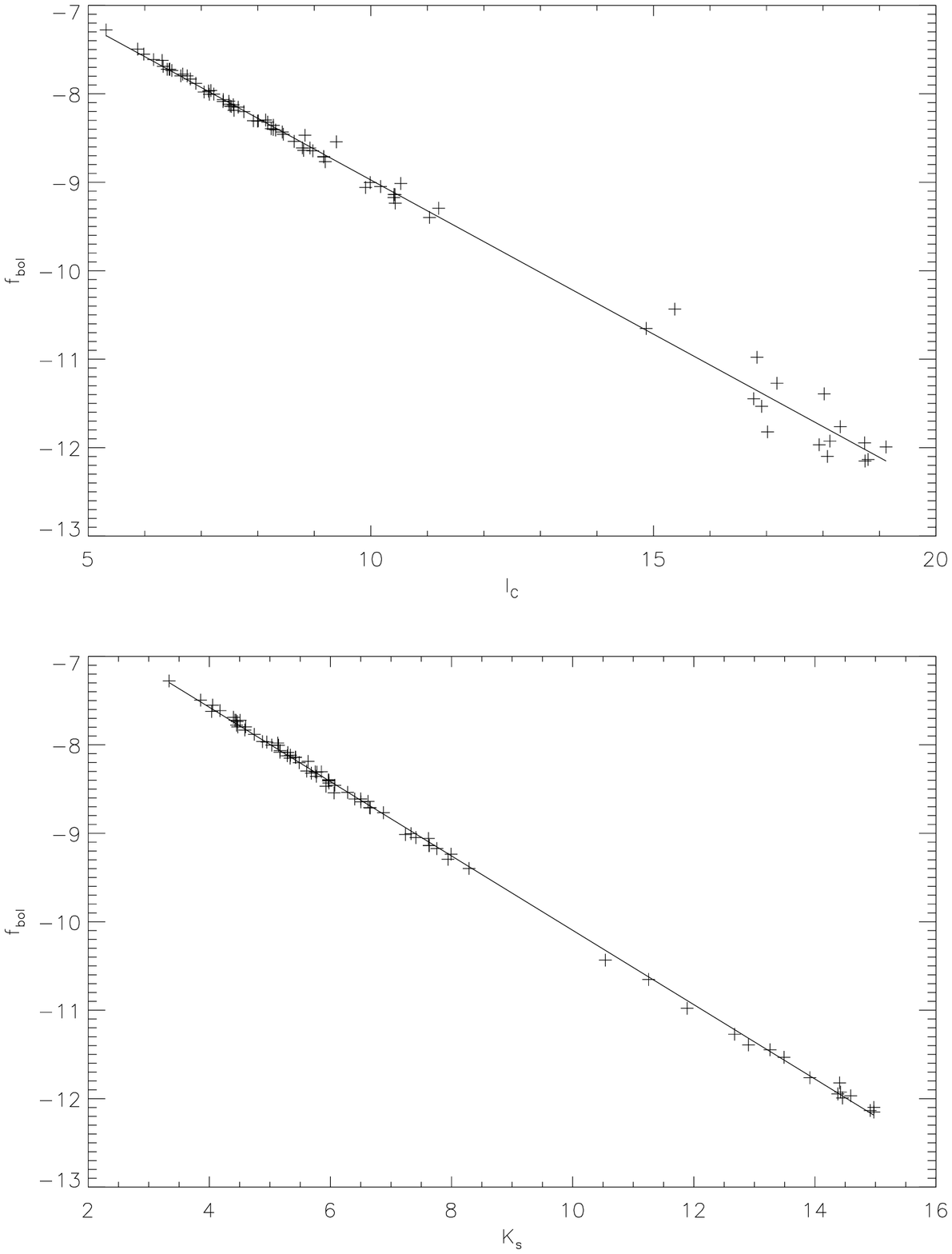}
\caption{f$_{\rm{bol}}$ for our sample as a function of I$_c$ and K$_s$.}
\label{fig4}
\end{center}
\end{figure}

\begin{figure}
\begin{center}
\includegraphics[width=0.7\textwidth]{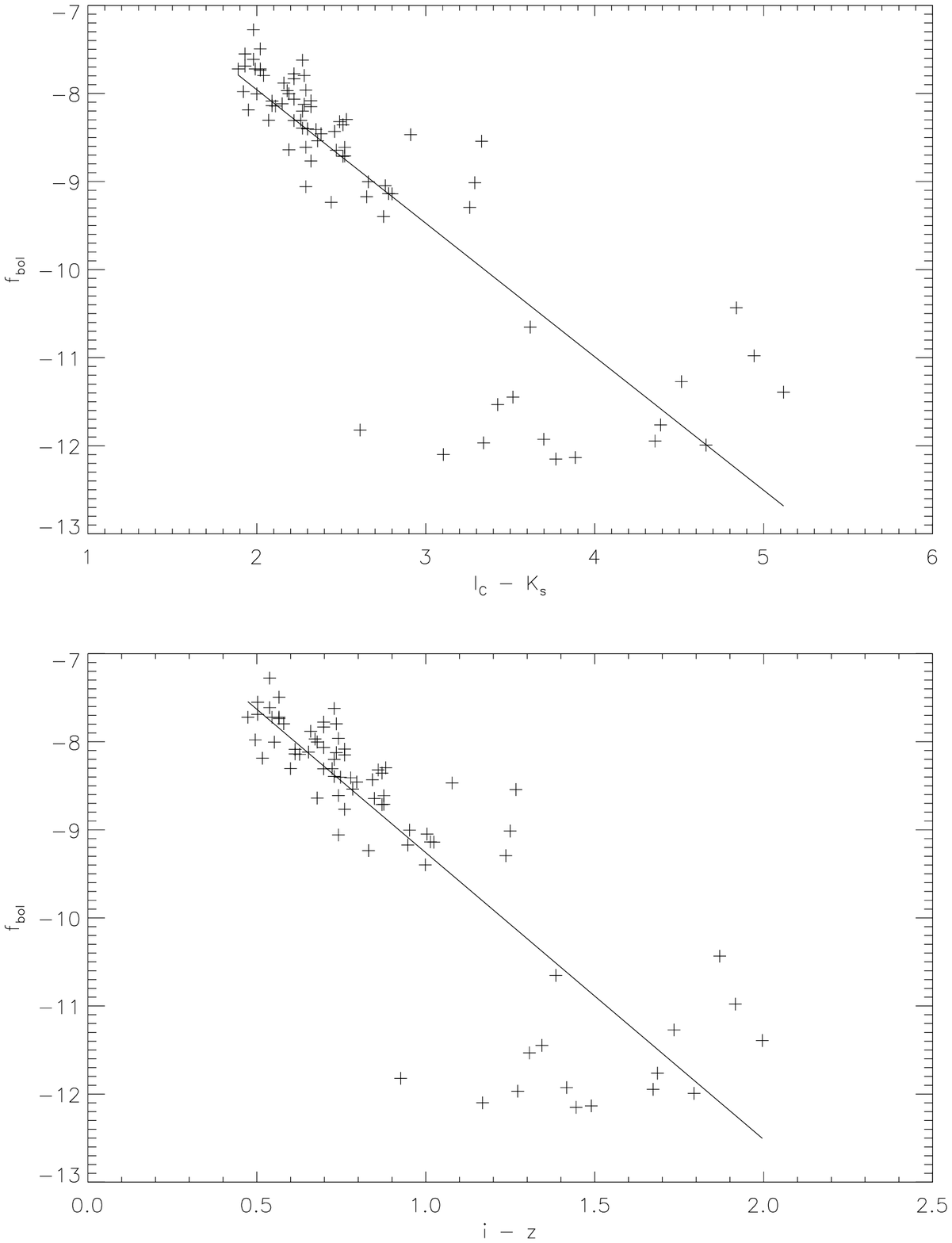}
\caption{f$_{\rm{bol}}$ for our sample as a function of I$_c$ - K$_s$ and $i - z$.}
\label{fig5}
\end{center}
\end{figure}

Equations 7 through 10 and Figures \ref{fig4} and \ref{fig5} provide fits to f$_{\rm{bol}}$ in terms of I$_C$, K$_s$, I$_C$ - K$_s$ and $i - z$, for use by anyone wishing to obtain, for example, L$_X$/L$_{\rm{bol}}$ for a large sample of stars.

\section{Results}

After having obtained $\chi$ across the spectral range, we computed the best fit to the distribution as a function of color. Fits to $\log(\chi)$ in V$-$I$_C$ and I$_C-$K$_s$ are presented below. Mean values for $\chi$ were computed for each spectral type, as shown in Table \ref{tbl-1}, enabling an estimate of an object's $\chi$ value in the event color data are not readily available or cannot be synthesized. 

The relationship between $\log(\chi)$ and color can be expressed by the polynomials:

\setcounter{equation}{0}
\begin{eqnarray}
\log(\chi) & = & -5.73342  +  3.07439(I_C-K_s) - 1.58615(I_C-K_s)^2  \nonumber\\
& & + 0.274372(I_C-K_s)^3 - 0.0154537(I_C-K_s)^4, \\
\nonumber\\
\log(\chi) & = & 0.607632 - 5.65757(V-I_C) + 2.55443(V-I_C)^2
\nonumber\\
& & - 0.500186(V-I_C)^3 + 0.0328053(V-I_C)^4 \\
\nonumber
\end{eqnarray}
\noindent
for ranges in color 1.89 $\leq$ I$_C-K_s$ $\leq$ 5.12, and 1.91 $\leq$ V$-I_C$ $\leq$ 4.40, respectively. 

Obtaining fits in terms of SDSS filters was somewhat more complicated. We computed transformations between traditional and SDSS colors using average colors by spectral type. The fits between SDSS and Cousins-2MASS colors, $r - i$ to V$-I_C$ and $i - z$ to I$_C-K_s$, are shown in Figure 2. The fits obtained are:
\begin{eqnarray}
i - z & = & -1.64309 + 1.59859(I_C-K_s) - 0.300041(I_C-K_s)^2 
\nonumber\\
& & + 0.0247425(I_C-K_s)^3,\\
\nonumber
r - i & = & 4.14067 - 4.52318(V-I_C) + 1.91395(V-I_C)^2 
\nonumber\\
& & - 0.225000(V-I_C)^3 \\
\nonumber
\end{eqnarray}

\noindent

Using these relations, we transformed our sample onto the SDSS system, and obtained fits to $\log(\chi)$ in terms of $r-i$ and $i-z$:
\begin{eqnarray}
\log(\chi) & = & -3.44258 - 0.509421(r-i)  \\
\log(\chi) & = & -3.31740 - 1.153440(i-z)\\
\nonumber
\end{eqnarray}
for ranges in color 0.92 $\leq$ $r-i$ $\leq$ 2.27, and 0.47 $\leq$ $i-z$ $\leq$ 2.00, respectively. 

In addition, we also provide fits to f$_{\rm{bol}}$ as a function of I$_C$, K$_s$, I$_C$ - K$_s$ and $i-z$ for our full sample:
\begin{eqnarray}
\log(f_{\rm{bol}}) & = & -5.48852 --0.34855 I_C\\
\log(f_{\rm{bol}}) & = & -5.89394 --0.42029 K_s\\
\log(f_{\rm{bol}}) & = & -4.92702 - 1.51571(I_C-K_s) \\
\log(f_{\rm{bol}}) & = & -6.00382 - 3.25601(i - z).\\
\nonumber
\end{eqnarray}

\section{Summary}

In order to study the strength of magnetic activity as a function of spectral type, we require a distance-independent way of determining the ratio L$_{\rm{H}\alpha}$/L$_{\rm{bol}}$. We used a sample of nearby stars and cool standards from 2MASS, spanning spectral types M0.5 to L0, with ranges in color 1.89 $\leq$ I$_C-K_s$ $\leq$ 5.12, and 1.91 $\leq$ V$-I_C$ $\leq$ 4.40. For the 8 parsec sample stars, we obtained the bolometric flux from photometry combined with bolometric corrections. For the 2MASS standards, optical photometry was obtained by convolving the spectra with filter response, then bolometric corrections were applied. The flux in the continuum near H$\alpha$ was obtained directly from the spectra. In both cases, we computed SDSS photometry for the sample using color transformations, provided in Section 3. We then found the ratio $\chi$ = f$_{\lambda6560}$/f$_{\rm{bol}}$. In Section 3, we provided fits to $\chi$ as a function of color (V - I$_C$, I$_C$ - K$_s$, $r - i$, $i - z$). Table 1 gives average $\chi$ values at each spectral type in the event photometry is not available, while Table 2 provides $\chi$ values for all stars in our sample. $\chi$ may be multiplied by the measured equivalent width in H$\alpha$ for a given star to yield the activity strength ratio, L$_{\rm{H}\alpha}$/L$_{\rm{bol}}$. If the reader possesses excellent quality data,  the method described here may also be used to calculate f$_{\lambda6560}$/f$_{\rm{bol}}$ for individual program stars.

This method has already yielded cleaner results for activity strength calculations in \citet{we04}, better illuminating trends in magnetic behavior at the end of the Main Sequence. Improved color transformations and a larger sample of stars with well-calibrated photometry and spectra are required to reduce the scatter in these relations.

\acknowledgements
We thank I. Neill Reid for providing data for the 8pc sample and late-type objects from 2MASS. This publication makes use of data products from the Two Micron All Sky Survey, which is a joint project of the University of Massachusetts and the Infrared Processing and Analysis Center/California Institute of Technology, funded by the National Aeronautics and Space Administration and the National Science Foundation.

\clearpage

\begin{deluxetable}{rrrrrrr}
\tabletypesize{\scriptsize}
\tablecaption{Mean $\chi$ by spectral type and average colors. \label{tbl-1}}
\tablewidth{0pt}
\tablehead{
\colhead{N} & \colhead{SpT}  & \colhead{V - I$_C$} & \colhead{I$_C$ - K$_s$} & \colhead{r-i} & \colhead{i-z} & \colhead{$\log(\chi)$} 
}
\startdata
4 &M0.5 V& 2.01 & 1.97 & \nodata &\nodata &$-$3.93619\\
4 &M1.0 V& 2.05 & 2.02 & 0.99 &0.50 &$-$3.93438 \\
5 &M1.5 V& 2.11 & 2.01 & \nodata &\nodata &$-$3.95412 \\
2 &M2.0 V& 2.20 & 2.04  & 1.09 &0.62 &$-$4.01516 \\
2 &M2.5 V& 2.36 & 2.14  &\nodata &\nodata &$-$4.11041 \\
7 &M3.0 V& 2.52 & 2.19  & 1.29 &0.73& $-$4.13192   \\
13&M3.5 V& 2.69  & 2.28 & \nodata &\nodata & $-$4.08665  \\
8 &M4.0 V& 2.84 & 2.39 & 1.57 &0.87&$-$4.19592  \\
9 &M4.5 V& 3.06  & 2.51 & \nodata &\nodata &$-$4.15220  \\
3 &M5.0 V& 3.46 & 3.15 & 1.98 &1.09 &$-$4.56243  \\
3 &M5.5 V& 3.61 & 2.87 & \nodata& \nodata &$-$4.46623  \\
3 &M6.0 V& 3.92 & 3.34 & 2.27  &1.27 &$-$4.75570 \\
3 &M6.5 V& 4.33 & 3.33 & \nodata& \nodata & $-$4.99468  \\
2 &M7.0 V& \nodata& 4.04 & 2.67 &1.52 &$-$5.28066 \\
2 &M7.5 V& \nodata & 3.43 & \nodata&\nodata &$-$5.01218 \\
2 &M8.0 V& \nodata & 3.99 & 2.82 &1.62 &$-$5.21965 \\
1 &M9.0 V& \nodata  & 4.66 & 2.89 &1.79 &$-$5.41719  \\
2 &M9.5 V& \nodata & 4.89 & \nodata &\nodata  &$-$5.36094 \\
2 &L0.0 V& \nodata  & 4.82 & 2.64 &1.85 &$-$5.32938 \\
 \enddata

\end{deluxetable}

\begin{deluxetable}{lrrrrr}
\tabletypesize{\scriptsize}
\tablewidth{0pt}
\tablenum{2}
\tablecaption{Full Sample Information \label{tbl-2}}
\tablehead{\colhead{Name\tablenotemark{a}
} & \colhead{SpT} & \colhead{V$-$I$_C$} & \colhead{I$_C$$-$K$_s$} & \colhead{BC$_K$} & \colhead{$\log(\chi)$} } 
\startdata
GJ 229 & M0.5 V & 2.01 & 1.98 & 2.49 & $-$3.998 \\
GJ 412 A & M0.5 V& 2.02 & 2.02 & 2.51 & $-$4.027 \\
GJ 514 & M0.5 V& 2.01 & 2.00 & 2.50 & $-$3.945 \\
GJ 809 & M0.5 V& 1.99 & 1.89 & 2.43 & $-$3.808 \\
GJ 15 A & M1.0 V& 2.14 & 1.93 & 2.46 & $-$3.855 \\
GJ 570 B & M1.0 V& 1.91 & 2.27 & 2.65 & $-$4.251 \\
GJ 686 & M1.0 V& 2.11 & 1.95 & 2.47 & $-$4.143 \\
GJ 908 & M1.0 V& 2.03 & 1.92 & 2.45 & $-$3.704 \\
GJ 205 & M1.5 V& 2.08 & 2.02 & 2.51 & $-$3.936 \\
GJ 526 & M1.5 V& 2.04 & 1.99 & 2.49 & $-$3.846 \\
GJ 625 & M1.5 V& 2.21 & 2.07 & 2.54 & $-$3.924 \\
GJ 880 & M1.5 V& 2.11 & 2.04 & 2.52 & $-$3.855 \\
GJ 1 & M1.5 V& 2.12 & 1.93 & 2.46 & $-$4.415 \\
GJ 393 & M2.0 V& 2.24 & 2.09 & 2.55 & $-$4.198 \\
GJ 411 & M2.0 V& 2.15 & 1.98 & 2.49 & $-$3.887 \\
GJ 408 & M2.5 V& 2.39 & 2.11 & 2.56 & $-$4.121 \\
GJ 752 A & M2.5 V& 2.32 & 2.16 & 2.59 & $-$4.100 \\
GJ 109 & M3.0 V& 2.45 & 2.26 & 2.64 & $-$4.024 \\
GJ 251 & M3.0 V& 2.53 & 2.22 & 2.62 & $-$4.165 \\
GJ 388 & M3.0 V& 2.51 & 2.22 & 2.62 & $-$4.077 \\
GJ 581 & M3.0 V& 2.51 & 2.22 & 2.62 & $-$4.228 \\
GJ 687 & M3.0 V& 2.50 & 2.22 & 2.62 & $-$4.479 \\
GJ 725 A & M3.0 V& 2.46 & 2.02 & 2.51 & $-$3.993 \\
GJ 860 A & M3.0 V& 2.68 & 2.19 & 2.61 & $-$4.120 \\
GJ 15 B & M3.5 V& 2.82 & 2.30 & 2.66 & $-$4.122 \\
GJ 190 & M3.5 V& 2.64 & 2.35 & 2.69 & $-$4.183 \\
GJ 273 & M3.5 V& 2.71 & 2.29 & 2.66 & $-$4.114 \\
GJ 445 & M3.5 V& 2.64 & 2.27 & 2.65 & $-$3.969 \\
GJ 628 & M3.5 V& 2.68 & 2.32 & 2.67 & $-$4.427 \\
GJ 643 & M3.5 V& 2.73 & 2.32 & 2.67 & $-$4.154 \\
G 203-047 A & M3.5 V& 2.81 & 2.47 & 2.74 & $-$4.584 \\
GJ 661 A & M3.5 V& 2.51 & 2.09 & 2.55 & $-$3.726 \\
GJ 725 B & M3.5 V& 2.55 & 2.18 & 2.60 & $-$4.092 \\
GJ 729 & M3.5 V& 2.78 & 2.32 & 2.67 & $-$4.012 \\
GJ 829 B & M3.5 V& 2.58 & 2.27 & 2.65 & $-$4.132 \\
GJ 873 & M3.5 V& 2.69 & 2.28 & 2.65 & $-$4.009 \\
GJ 896 A & M3.5 V& 2.84 & 2.15 & 2.59 & $-$4.168 \\
GJ 1005 B & M4.0 V& 2.77 & 2.19 & 2.61 & $-$4.019 \\
GJ 105 B & M4.0 V& 2.78 & 2.29 & 2.66 & $-$4.039 \\
GJ 169.1 A & M4.0 V& 2.81 & 2.51 & 2.76 & $-$4.323 \\
GJ 213 & M4.0 V& 2.82 & 2.36 & 2.69 & $-$4.338 \\
GJ 300 & M4.0 V& 2.90 & 2.51 & 2.76 & $-$4.311 \\
GJ 447 & M4.0 V& 2.98 & 2.53 & 2.76 & $-$4.067 \\
GJ 555 & M4.0 V& 2.86 & 2.46 & 2.74 & $-$4.571 \\
GJ 699 & M4.0 V& 2.78 & 2.28 & 2.65 & $-$4.171 \\
GJ 54.1 & M4.5 V& 3.13 & 2.52 & 2.76 & $-$4.149 \\
GJ 83.1 & M4.5 V& 3.12 & 2.52 & 2.76 & $-$4.130 \\
GJ 166 C & M4.5 V& 2.87 & 2.38 & 2.70 & $-$4.034 \\
GJ 285 & M4.5 V& 2.95 & 2.49 & 2.75 & $-$3.981 \\
GJ 299 & M4.5 V& 2.92 & 2.29 & 2.66 & $-$4.422 \\
GJ 1224 & M4.5 V& 3.18 & 2.44 & 2.73 & $-$4.066 \\
LHS 3376 & M4.5 V& 3.12 & 2.65 & 2.81 & $-$4.196 \\
LHS 3799 & M4.5 V& 3.22 & 2.66 & 2.81 & $-$4.345 \\
2MJ2300189$+$121024 & M4.5 V& \nodata & 2.61 & 2.80 & $-$4.226 \\
GJ 1156 & M5.0 V& 3.46 & 2.78 & 2.85 & $-$4.632 \\
GJ 905 & M5.0 V& 3.45 & 2.91 & 2.88 & $-$4.387 \\
2MJ0244463$+$153531A & M5.0 V& \nodata & 3.77 & 3.06 & $-$4.749 \\
GJ 1002 & M5.5 V& 3.59 & 2.76 & 2.84 & $-$4.419 \\
GJ 1286 & M5.5 V& 3.63 & 2.75 & 2.84 & $-$4.449 \\
2MJ0244463$+$153531B  & M5.5 V& \nodata & 3.10 & 2.93 & $-$4.539 \\
GJ 406 & M6.0 V& 4.06 & 3.33 & 2.93 & $-$4.925 \\
GJ 412 B & M6.0 V& 3.77 & 2.8 & 2.85 & $-$4.562 \\
2MJ0435490$+$153720 & M6.0 V& \nodata & 3.89 & 3.07 & $-$4.874 \\
GJ 1111 & M6.5 V& 4.26 & 3.29 & 2.93 & $-$5.022 \\
LHS 292 & M6.5 V& 4.40 & 3.26 & 2.93 & $-$5.089 \\
2MJ0242252$+$134313 & M6.5 V& \nodata & 3.43 & 2.99 & $-$4.896 \\
2MJ0055584$+$275652 & M7.0 V& \nodata & 3.70 & 3.04 & $-$5.238 \\
2MJ0855559$+$380343 & M7.0 V& \nodata & 4.39 & 3.14 & $-$5.328 \\
2MJ0348036$+$234411 & M7.5 V& \nodata & 3.52 & 3.01 & $-$4.973 \\
2MJ2258590$+$152047 & M7.5 V& \nodata & 3.34 & 2.98 & $-$5.055 \\
2MJ1047138$+$402649 & M8.0 V& \nodata & 3.62 & 3.03 & $-$5.037 \\
2MJ1434264$+$194050 & M8.0 V& \nodata & 4.36 & 3.14 & $-$5.542 \\
2MJ0251222$+$252124 & M9.0 V& \nodata & 4.66 & 3.18 & $-$5.417 \\
2MJ0024246$-$015820 & M9.5 V& \nodata & 4.84 & 3.20 & $-$5.292 \\
2MJ1733189$+$463400 & M9.5 V& \nodata & 4.94 & 3.21 & $-$5.442 \\
2MJ0345432$+$254023  & L0.0 V& \nodata & 4.51 & 3.16 & $-$5.312 \\
2MJ0058425$-$065124 & L0.0 V& \nodata & 5.12 & 3.23 & $-$5.347 \\
\enddata
\tablenotetext{a}{2M: Cutri et al., 2MASS All-Sky Catalog of Point Sources, 2003.} 
\end{deluxetable}

\clearpage

\pagestyle{empty}

\end{document}